\documentclass[conference]{IEEEtran}
\IEEEoverridecommandlockouts
\usepackage{cite}
\usepackage{amsmath,amssymb,amsfonts}
\usepackage{algorithmic}
\usepackage{algorithm}
\usepackage{graphicx}
\usepackage{textcomp}
\usepackage{xcolor}
\usepackage{booktabs}
\usepackage{multirow}
\newcommand{\mat}[1]{\boldsymbol{#1}}

\usepackage{booktabs}
\usepackage{graphicx}
\def\BibTeX{{\rm B\kern-.05em{\sc i\kern-.025em b}\kern-.08em
    T\kern-.1667em\lower.7ex\hbox{E}\kern-.125emX}}
\begin{document}

\title{Integrated Simulation Framework for Adversarial Attacks on Autonomous Vehicles\\


\thanks{This work has received funding from the European Union’s Horizon Europe research and innovation programme under the GuardAI project, “Robust and Secure Edge AI Systems for SafetyCritical Applications” (Grant Agreement No. 101168067).
}
}

\author{
\IEEEauthorblockN{Christos Anagnostopoulos$^{2,3}$, Ioulia Kapsali$^{1}$, Alexandros Gkillas$^{1,2}$, Nikos Piperigkos$^{1,2}$,  Aris S. Lalos$^{1}$}
\IEEEauthorblockA{$^1$Industrial Systems Institute, Athena Research Center, Patras Science Park, Greece\\
$^2$AviSense.AI, Patras Science Park, Greece\\
$^3$ Dpt. of Informatics \& Telecom., University of Ioannina, Arta, Greece\\
Emails: \{anagnostopoulos, gkillas, piperigkos\}@avisense.ai, \{ioulia.kapsali, lalos\}@athenarc.gr
}
}

\maketitle

\begin{abstract}
Autonomous vehicles (AVs) rely on complex perception and communication systems, making them vulnerable to adversarial attacks that can compromise safety. While simulation offers a scalable and safe environment for robustness testing, existing frameworks typically lack comprehensive support for modeling multi-domain adversarial scenarios. This paper introduces a novel, open-source integrated simulation framework designed to generate adversarial attacks targeting both perception and communication layers of AVs. The framework provides high-fidelity modeling of physical environments, traffic dynamics, and V2X networking, orchestrating these components through a unified core that synchronizes multiple simulators based on a single configuration file. Our implementation supports diverse perception-level attacks on LiDAR sensor data, along with communication-level threats such as V2X message manipulation and GPS spoofing. Furthermore, ROS 2 integration ensures seamless compatibility with third-party AV software stacks. We demonstrate the framework's effectiveness by evaluating the impact of generated adversarial scenarios on a state-of-the-art 3D object detector, revealing significant performance degradation under realistic conditions.
\end{abstract}


\section{Introduction}
Autonomous vehicles have the potential to revolutionize transportation, promising enhanced safety, efficiency, and accessibility, with sophisticated perception and communication systems central to their operation. However, their increasing complexity and connectivity also make AVs vulnerable to adversarial attacks. Malicious actors can subtly manipulate sensor inputs, particularly for 3D object detectors crucial for interpreting sensor data, or communication signals, potentially causing catastrophic failures and severely degrading accuracy and reliability \cite{huangV2XCooperativePerception2024}. Real-world testing of edge cases, such as sensor malfunctions, severe weather, or intentional attacks, is costly, time-consuming, and poses significant safety and legal challenges.\\ 
Consequently, high-fidelity simulators \cite{liAdvGPSAdversarialGPS2024a,ramakrishnaANTICARLAAdversarialTesting2022a,finkenzellerSimutackAttackSimulation2023,xiangV2XPASGGeneratingAdversarial2023} have emerged as essential tools, enabling rigorous, repeatable, and systematic evaluation of AV performance under controlled adversarial conditions, thus offering a scalable and safe method for assessing and improving AV robustness before real-world deployment. To this end, Xiang et al. \cite{xiangV2XPASGGeneratingAdversarial2023} proposed V2XP-ASG, an adversarial scene generator for LiDAR-based multi-agent perception, to identify challenging corner cases. In a similar context, ANTI-CARLA \cite{ramakrishnaANTICARLAAdversarialTesting2022a} provides a framework specifically designed to generate adversarial test cases aimed at causing failures in autonomous vehicle systems within the CARLA \cite{dosovitskiyCARLAOpenUrban2017} simulator environment. However, both of these works concentrate on attack scenario generation rather than delivering a fully integrated simulation environment. In contrast, Simutack \cite{finkenzellerSimutackAttackSimulation2023} offers an open-source attack simulation framework that supports various attack surfaces but lacks support for LiDAR-based attacks and does not provide integration with the ROS 2 \cite{ros2} ecosystem.\\
In contrast to the aforementioned solutions, our proposed framework offers a comprehensive, modular, and extensible simulation environment specifically designed to generate adversarial attacks targeting LiDAR point clouds, alongside communication-layer threats such as V2X message manipulation and GPS spoofing. By supporting coordinated attacks across both perception and communication surfaces, the framework enables the study of complex, multi-domain adversarial scenario. In addition, it is built with ROS 2 integration, ensuring seamless interoperability with third-party AV stacks while supporting enhanced security, modularity, and quality of service (QoS).\\
The main contributions of this work can be summarized as follows:
\begin{itemize}
\item \textbf{Novel Simulation Framework for Adversarial Attacks:} It has been developed a new simulation framework capable of generating adversarial data at both the 3D perception and communication levels in autonomous driving scenarios. The framework takes as input a single configuration file, which defines the simulation parameters, scenario layout, and attack settings. Based on this file, it automatically initializes and synchronizes all simulators, configures sensor suites, and activates specific adversarial behaviors.    
    \item \textbf{ROS 2 Integration:} The proposed work offers ROS 2 compatibility, facilitating seamless integration with third-party autonomous driving software.   
    \item \textbf{SOTA Evaluation:} It is provided and evaluation of the effectiveness and validity of the generated adversarial data using a representative use case, demonstrating the practical impact of the proposed framework in realistic settings.
\end{itemize}

\section{System}
\begin{figure}
\centering
 \includegraphics[scale=0.25]{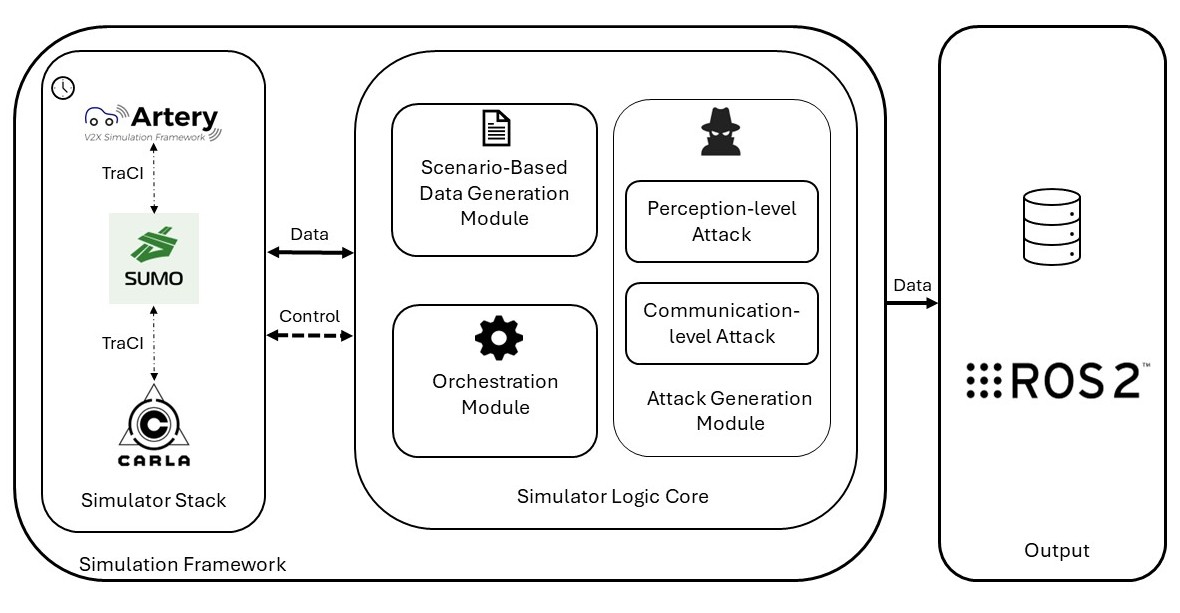}
\caption{Overview of the proposed integrated simulation framework. On the left, the Simulator Stack contains CARLA, SUMO, and Artery, interconnected via the TraCI protocol. Artery acts as the master clock, orchestrating synchronized progression across the simulation components, with SUMO and CARLA operating as synchronized slaves. The central Simulation Logic Core module comprises three key submodules. The \textit{Scenario-Based Data Generation Module} parses a single scenario configuration file and initializes all simulators with appropriate parameters. The \textit{Orchestration Module} manages simulation synchronization under two modes: (i) in SUMO-driven scenarios, SUMO acts as the primary traffic simulator, defining and controlling the vehicles. The Orchestration Module translates SUMO’s vehicle states into corresponding entities in CARLA (ii) in CARLA-driven scenarios, the Orchestration Module controls vehicle spawning in CARLA and pushes vehicle states into SUMO. The \textit{Attack Generation Module} supports both perception-level and communication-level adversarial attacks. Finally, the system either stores the produced data in common formats or publishes it as ROS 2-compatible messages for downstream processing.}

  \label{fig:arghitecture}
\end{figure}
Our proposed system builds upon the work presented in \cite{anagnostopoulosOpenSourceIntegratedSimulation2022}, extending it by introducing a new orchestration pipeline that supports simulation scenarios based on either traffic scenario descriptions via SUMO \cite{SUMO} or autonomous vehicle environments via CARLA. An overview of the architectural elements of the framework is presented in Figure~\ref{fig:arghitecture}. The architecture integrates a simulator stack comprising the three aforementioned simulators with a Simulation Logic Core (SLC) module responsible for initializing, orchestrating, and recording each simulation session.

Specifically, the SLC consists of three key components: a scenario-based data generation module for initializing the simulators based on a unified scenario description file, an orchestration module for synchronizing and controlling the simulators, and an attack generation module that includes both a 3D perception attack engine capable of injecting adversarial point cloud data and a communication-level adversarial attack engine targeting the V2X layer. Finally, the framework is designed to integrate seamlessly with ROS 2, supporting Quality of Service configurations and meeting relevant security requirements.
\subsection{Overview}The proposed simulator offers a modular, open-source framework that integrates CARLA, SUMO, and Artery/OMNeT++ \cite{ArteryV2XSimulation} to support high-fidelity simulation of cooperative autonomous vehicle (CAV) systems. By combining physical dynamics, traffic flow, and V2X communication layers, the platform enables end-to-end testing of perception, networking, and decision-making modules using unaltered ROS-compatible code.
CARLA is used for rendering and simulating the physical environment, including realistic sensor data (LiDAR, cameras, radar). Through the CARLA-ROS bridge \cite{CarlasimulatorRosbridge2025}, this data is streamed into the ROS ecosystem, allowing for convenient and modular algorithm testing. SUMO handles traffic simulation and acts as the central synchronization controller via its TraCI interface, coordinating simulation steps across all subsystems.
Artery, built on OMNeT++, simulates V2X communication using ETSI ITS-G5 protocols, which is a family of standards for implemening relevant communications. It exchanges data with SUMO to track vehicle mobility and directly to the main synchronization module which amongst others publishes Cooperative Awareness Messages (CAMs) to ROS enabling the testing of cooperative behaviors such as multi-agent localization and fusion.

\subsection{Orchestration Module}
This module is responsible for synchronizing the three different simulators and aligning their operation with the objectives defined in the initial scenario description. In the proposed setup, Artery functions as the master clock, while SUMO and CARLA operate as synchronized slaves. Two distinct use cases have been implemented to support different simulation workflows:

\begin{itemize}
    \item \textbf{SUMO-driven scenarios:} SUMO acts as the primary traffic simulator, defining and controlling vehicle behavior. The Orchestration Module translate SUMO’s vehicle states into corresponding entities in CARLA    
    \item \textbf{CARLA-driven scenarios:} The Orchestration Module manages vehicles spawning and behavior in CARLA. The resulting vehicle states are pushed into SUMO to maintain synchronization. 
\end{itemize}
In both cases, the module is responsible for the smooth and synchronized operation of the simulation, as well as for coordinating data acquisition from the different sources. To this end, it maintains a direct connection with Artery for acquiring V2X communication data as it is generated by the simulator, ensuring consistent temporal alignment between all integrated components.
\subsection{Scenario-Based Data Generation Module}
This component is responsible for parsing an initial scenario description file and configuring the simulators accordingly. The file includes both general simulation parameters, such as duration and time step, and simulator-specific settings. These settings include weather conditions, sensor configurations, and number of vehicles for CARLA, traffic routes and vehicle behavior for SUMO, and transmission power and emission parameters for Artery.\\
In addition, it gathers synchronized data streams from the various sensors deployed in the simulation environment. The sensor data are either stored locally in commonly used formats or, through the CARLA ROS 2 bridge (as described in \ref{ros_integration}), transformed and made available in ROS-compatible formats
\begin{algorithm}
\caption{Simulation Synchronization}
\begin{algorithmic}[1]
\renewcommand{\algorithmicrequire}{\textbf{Input:}}
\renewcommand{\algorithmicensure}{\textbf{Output:}}
\REQUIRE Scenario configuration, Artery clock step $\Delta t$
\ENSURE  Simulated Dataset
\\ \textit{Initialization:}
\STATE Launch Artery, SUMO, CARLA, and SLC
\IF {Scenario mode is SUMO-defined}
    \STATE Load SUMO network and routes
    \STATE Synchronize CARLA with SUMO and CARLA acts as follower
\ELSE
    \STATE Spawn actors in CARLA
    \STATE Synchronize SUMO with CARLA and SUMO acts as follower
\ENDIF
\STATE Establish TraCI connection (Artery $\leftrightarrow$ SUMO)
\STATE Orchestration module opens a socket connection to Artery
\\ \textit{Main Loop:}
\WHILE {simulation is running}
    \STATE Artery calls \texttt{simulateUntil($t + \Delta t$)}
    \STATE Artery steps SUMO
    \IF {Scenario mode is SUMO-defined}
        \STATE Read SUMO actor states via TraCI
        \STATE Update CARLA actor states accordingly
    \ELSE
        \STATE Read CARLA actor states
        \STATE Update SUMO actor states accordingly
    \ENDIF
    \STATE Artery updates all actor states
    \STATE Artery generates and emits CAMs
\STATE Sensor data streams from all simulators are stored or converted into ROS 2 messages

\ENDWHILE
\RETURN Generated dataset
\end{algorithmic}
\end{algorithm}

\subsection{Attack Generation Module}
This component supports both perception-level and communication-level adversarial attacks. The implemented perception-level attacks, specifically targeting point clouds, are described in detail in Section~\ref{perception attacks}. These include perturbation, detachment, and attachment of points. The module initializes the adversarial point cloud using the original data generated by CARLA. Depending on the selected attack scenario, it either perturbs existing points to maximize a loss function under specific constraints, adds new adversarial points, or selectively removes critical points. 
Regarding the communication layer, the module implements a series of well-established adversarial strategies and GPS spoofing attacks, as detailed in Section~\ref{com attacks}.

\subsection{ROS 2 Integration} 
\label{ros_integration}
The simulation framework is implemented to integrate seamlessly with the ROS 2 middleware, providing an efficient environment for algorithm testing. It extends the CARLA ROS 2 bridge \cite{CarlasimulatorRosbridge2025} and incorporates a virtual V2X sensor which has a dual purpose. Regarding V2X communication, the framework collects and publishes Cooperative Awareness Messages that are compliant with the ETSI ITS standards. Additionally, it publishes the Local Dynamic Map (LDM) for each vehicle. \\
The Local Dynamic Map encapsulates information about an agent's surrounding environment. It is a conceptual data structure within each Intelligent Transport System (ITS), defined under the V2X communication paradigm as described in ETSI EN 302 665 standard. The LDM enhances the perception capabilities of participating actors by providing contextual information about nearby entities. In this simulation framework, the LDM is published using ROS 2 messages to facilitate interoperability.

\section{Attack Model Design and Implementation}
\subsection{Perception-Level Adversarial Attack} \label{perception attacks}

Perception-level adversarial attacks operate by directly manipulating the digital input to a LiDAR-based 3D object detector, without requiring any modifications to the physical environment. These attacks can be broadly categorized into \textit{point perturbation, point detachment and point attachment} \cite{zhang2024comprehensive}. Let the input point cloud be defined as \( \mat{P} = \{\mat{p}_i \in \mathbb{R}^3 \mid i = 1, ..., N\} \), where each point \( \mat{p}_i = (x_i, y_i, z_i) \) corresponds to a 3D coordinate acquired by a LiDAR sensor. The annotated 3D ground-truth bounding boxes are denoted as \( \mathcal{G} = \left\{ \mathbf{B}_j \in \mathbb{R}^8 \right\} \), where each bounding box \( \mathbf{B}_j = (x_j, y_j, z_j, l_j, w_j, h_j, r_j, c_j) \) encodes the object’s center position, dimensions, orientation angle, and class label.

\subsubsection{Adversarial Point Perturbation}

Adversarial point perturbation attacks aim to deceive 3D object detectors by introducing small but structured modifications to the 3D coordinates of 
points in a point cloud. Let \( \boldsymbol{\delta} = \{ \boldsymbol{\delta}_i \in \mathbb{R}^3 \mid i = 1, \dots, N \} \) represent the perturbations applied to each point. The adversarial point cloud is then given by \( \mat{P}_a = \{ \mathbf{p}_i^{\,a} = \mathbf{p}_i + \boldsymbol{\delta}_i \} \).
We consider an untargeted adversarial setting, where the attacker’s goal is to disrupt the detection pipeline without targeting a specific object class or location. 

The adversarial point cloud is crafted to degrade the performance of the detector while preserving geometric and visual similarity to the original input. This is formulated as an optimization problem with a dual-objective loss comprising:

\begin{equation}
\mathcal{L} = \mathcal{L}_{\text{adv}} + \lambda \cdot \mathcal{L}_{\text{per}}
\label{eq:opt}
\end{equation}

\noindent
where $\lambda$ a hyperparameter balancing the two objectives.

\textbf{Adversarial Loss} ($\mathcal{L}_{\text{adv}}$): Maximizes the detector’s prediction error, encouraging misclassification or localization errors. It is defined as: $\mathcal{L}_{\text{adv}} = -\mathcal{L}_{\text{det}}(\mat{P}_a, G)$,
where $\mathcal{L}_{\text{det}}$ is the detection loss.

\textbf{Perturbation Loss} ($\mathcal{L}_{\text{per}}$): Penalizes excessive deviation from the original geometry, constraining the magnitude of perturbations via the point-wise $\ell_2$ norm: $\mathcal{L}_{\text{per}} = \sum_{i=1}^{N} \| \mat{\delta}_i \|_2^2$,

To generate adversarial examples, we adopt a Projected Gradient Descent (PGD) \cite{PGD} scheme. At initialization, small random perturbations sampled uniformly from \( \mathcal{U}(-\epsilon, \epsilon) \) are added to each point. Then, at each iteration \( t \), the adversarial point cloud is updated via a normalized gradient step and projected back onto an \( \ell_2 \)-ball of radius \( \epsilon \) centered at the original point cloud \( P \), ensuring the perturbation remains bounded:
\begin{equation}
\mat{P}_a^{t+1} = \text{Clip}_{\mat{P}, \epsilon} \left( \mat{P}_a^t - \alpha \cdot \frac{\nabla_{\mat{P}_a^t} \mathcal{L}}{ \| \nabla_{\mat{P}_a^t} \mathcal{L} \|_2 } \right),
\end{equation}
where $\alpha$ is the step size, $\epsilon$ defines the maximum allowed displacement for each point and the clipping function $\text{Clip}_{\mat{P}, \epsilon}(\cdot)$ implements point-wise clipping of point cloud and is denoted as: 
\[
\text{Clip}_{\mat{P}, \epsilon}({\mat{P}_a}) = \left\{ \mathbf{p}_i^{\,a} \,\middle|\, 
\mathbf{p}_i^{\,a} = 
\begin{cases}
\mathbf{p}_i + \boldsymbol{\delta}_i, & \text{if } \|\boldsymbol{\delta}_i\|_2 \leq \epsilon \\
\mathbf{p}_i + \epsilon \cdot \dfrac{\boldsymbol{\delta}_i}{\|\boldsymbol{\delta}_i\|_2}, & \text{otherwise}
\end{cases}
\right\}
\]

\subsubsection{Adversarial Point Detachment}
 
To exploit the inherent vulnerability of LiDAR point clouds to sparsity, adversaries can selectively remove critical points to disrupt the performance of 3D object detectors. This attack produces an adversarial point cloud \( \mathcal{P}_a \subset \mathcal{P} \) by dropping a subset of strategically chosen points such that the detector’s output is significantly degraded. Following the approach introduced by \cite{Zheng_2019_ICCV}, the attack's effectiveness relies on identifying critical points that most influence the model’s predictions. At each point \( \mat{p}_i \) a saliency score \( s_i \) is defined as:
\begin{equation}
s_i = \left\| \nabla_{\mat{p}_i} \mathcal{L}_{\text{det}}(\mat{P}, G) \right\|_2,
\label{eq:saliency}
\end{equation}
where \( \nabla_{\mat{p}_i} \) denotes the gradient of the detection loss \( \mathcal{L}_{\text{det}} \) with respect to the point \( \mat{p}_i \). A higher score indicates greater importance in the model's decision-making process. The saliency map is represented as \( \mathcal{S} = \{ (\mat{p}_i, s_i) \} \), pairing each point with its corresponding importance score. 
The adversarial point cloud is generated by iteratively removing the most salient points, adopting a greedy strategy: after each removal, the saliency map is updated to reflect the modified input and its effect on the model.


\subsubsection{Adversarial Point Attachment}

This attack strategy aims to degrade the performance of 3D object detectors by enriching the input point cloud with a small number of deliberately placed synthetic points. Formally, the modified point cloud is defined as \( \mat{P}_a = \mat{P} \cup \mat{Z} \), where \( \mat{Z} = \{ \mat{z}_i \in \mathbb{R}^3 \mid i = 1, \dots, K \} \) represents the set of newly injected points. 

Rather than inserting points randomly, we adopt the initialize-and-shift approach~\cite{xiang2019generating}, which reduces the complexity of searching the entire 3D space for optimal insertion locations. Guided by the saliency scores computed via Eq.~(3), we identify the spatial regions most critical to the model’s predictions. Specifically, the top-\(K\) most salient points themselves are directly selected as the initial injected points, ensuring they lie in high-impact regions. After initialization, the placement of these injected points is refined by solving the optimization problem in Eq.\ref{eq:opt}, which balances adversarial effectiveness and spatial plausibility. To ensure the geometric similarity between the original point cloud and its adversarial counterpart, we incorporate the Chamfer Distance \( \mathcal{D}_C \) as a regularization constraint, defined as:
\begin{equation}
\begin{split}
D_C(\mat{P}, \mat{P}_a) = \; & \frac{1}{|\mat{P}|} \sum_{\mathbf{p}_i \in \mat{P}} \min_{\mathbf{q}_m \in \mat{P}_a} \| \mathbf{p}_i - \mathbf{q}_m \|_2^2 \\
+ & \frac{1}{|\mat{P}_a|} \sum_{\mathbf{q}_m \in \mat{P}_a} \min_{\mathbf{p}_i \in \mat{P}} \| \mathbf{q}_m - \mathbf{p}_i \|_2^2
\end{split}
\label{eq:chamfer}
\end{equation}
, where \( \mathbf{q}_m \in \mat{P}_a \) refers to a point in the adversarial point cloud. The number of inserted points is limited to maintain imperceptibility and preserve physical plausibility.

\subsection{Communication Level Attacks } \label{com attacks}

Cooperative communication between AVs can undoubtedly enhance overall perception accuracy and improve safety through shared situational awareness \cite{cooperative}. However, this connectivity also introduces new security risks. A malicious actor can forge, alter, or replay CAMs, thereby undermining the integrity of the shared information. For example, in a fake position or speed attack, the adversary manipulates the contents of CAMs by broadcasting falsified coordinates or velocities. A more advanced variant is the Sybil attack \cite{sybil}, where a single attacker impersonates multiple distinct vehicles by sending out multiple CAMs with different vehicle IDs and locations, effectively generating “ghost” vehicles within the network compromising the integrity of cooperativeness, potentially leading to unsafe behaviors. In our implementation using Artery, this attack is realized by instantiating multiple virtual ghost nodes, each configured to broadcast unique fake positions.\\
Beyond V2X message manipulation, the module also simulates GPS spoofing attacks, which aim to deceive a vehicle's navigation system by broadcasting counterfeit GPS signals. This can lead to incorrect localization, path planning errors, and potentially dangerous maneuvers.
\begin{itemize}
    \item \textbf{Random Bias Attack (RBA) \cite{Randombiasattack}}: RBA is a simple spoofing strategy where the attacker injects a random constant error (bias) into the communicated data. In our context, this means replacing a vehicle’s true GPS pose with a corrupted pose offset by a random bias vector. The bias is typically drawn from a uniform distribution within a certain range, chosen to reflect the maximum plausible GPS error or spoofing signal strength. Formally, one can model the attack as $x_{\text{adv}} = x + \delta$, where $\delta_i \sim \mathcal{U}[-\Delta_i, \Delta_i]$ for each dimension $i$ (with $\Delta_i$ set by allowable error bounds). This random bias attack does not use gradient information, but instead, it simulates a non-targeted perturbation such as natural GPS noise or unsophisticated jamming. Despite its simplicity, RBA serves as a baseline for GPS spoofing in simulation, representing scenarios where an attacker induces a bounded but unpredictable error in the vehicle-to-vehicle messages.
    \item \textbf{Position Altering Attack (PAA) \cite{positionalteringattack}}: PAA is a more aggressive communication-level attack in which an adversary falsifies the vehicle’s communicated position to a significantly incorrect value. Rather than a small perturbation, PAA involves spoofing the GPS message so that the victim perceives the vehicle at a fictitious location far from its true position. This could be achieved by injecting a large bias or entirely fabricated coordinates into the message stream. In simulation, PAA is implemented by altering the position data beyond normal error bounds, thus effectively simulating a strong GPS spoofer that relocates the vehicle’s reported position. The result is a communicated pose that is plausible but erroneous enough to severely mislead cooperative perception or planning, like for instance, causing a receiving vehicle to misalign sensor data. PAA exemplifies a worst-case communication attack, demonstrating how a determined adversary could compromise the coordination among AVs by delivering highly corrupted location messages.
    \end{itemize}

\section{System Evaluation}
\subsection{Metrics}
To evaluate the effectiveness and stealthiness of the proposed adversarial attacks, we employ two complementary metrics: one to assess the robustness of the detector, and one to quantify the perceptibility of the adversarial examples.

\paragraph{Adversarial Impact Assessment} We adopt a relative metric known as the \emph{mAP ratio} (\%), which is defined as the ratio of the mean Average Precision (mAP) computed on adversarial examples to that computed on clean point clouds over the entire validation set. Formally, let \( \text{mAP}_{\text{adv}} \) and \( \text{mAP}_{\text{clean}} \) denote the mAP values on adversarial and clean inputs, respectively. Then:
\[
\text{mAP ratio} = \frac{\text{mAP}_{\text{adv}}}{\text{mAP}_{\text{clean}}}
\]

 Lower values of mAP ratio correspond to stronger attacks that more severely impair detection accuracy, whereas higher values suggest greater model robustness and limited adversarial effectiveness.
\paragraph{Perceptibility Measurement}
It is expected that the perturbed point cloud remains visually and geometrically similar to the original input. To quantify this similarity, we use the Chamfer Distance (CD) as a perceptual metric. 
\subsection{Simulation Setup}
\paragraph{Dataset}
The dataset was collected in Town06 of CARLA over the course of 3,000 frames at a sampling rate of 10 Hz, corresponding to 5 minutes of simulated urban driving. The ego vehicle was equipped with a 64-channel LiDAR sensor, and in addition multiple other vehicles were spawned. The dataset includes raw LiDAR point clouds, ground-truth positions and orientations of all vehicles, and associated metadata such as timestamps and transformation matrices.
 \paragraph{Adversarial Attack Implementation}
 We use SECOND \cite{yanSECONDSparselyEmbedded2018}, a state-of-the-art 3D object detector, as the target model to evaluate the effectiveness of our adversarial attacks.
We adopt the PGD approach to generate adversarial examples for the point perturbation attack. For iterative attacks, each adversarial sample is iteratively updated over 40 steps, with a step size defined as \( \alpha = \epsilon / 30 \). For point perturbation attack, the perturbation budget \( \epsilon \) is evaluated over the set \{0.5, 1, 3, 5, 7, 10\}\,cm.
For point detachment attacks, we remove a fraction of the original points progressively across 10 iterations. The detachment ratio is selected from \{0.001, 0.003, 0.005, 0.007, 0.01, 0.015\} to explore the impact of different sparsity levels.
In the point attachment scenario, 300 artificial points are injected into the point cloud to deceive the detector. The injected points are further perturbed with $\ell_2$-norm constraints chosen from \{0.05, 0.1, 0.3, 0.5, 0.7, 1\}\,m, regulating their spatial displacement from their initial positions.

\begin{table}
\centering
\caption{Comparison of attack types on the second detector with Chamfer Distance (CD) and mAP ratio.}
\label{tab:attack_summary}
\resizebox{0.5\textwidth}{!}{%
\begin{tabular}{lcccccc}
\toprule
\textbf{Detector} & \multicolumn{2}{c}{\textbf{Perturbation}} & \multicolumn{2}{c}{\textbf{Detachment}} & \multicolumn{2}{c}{\textbf{Attachment}} \\
& \textbf{CD} & \textbf{mAP ratio} & \textbf{CD} & \textbf{mAP ratio} & \textbf{CD} & \textbf{mAP ratio} \\
\midrule
SECOND & 0.01533 & 73.20 & 0.07035 & 91.77 & 0.00126 & 78.46 \\
\bottomrule
\end{tabular}%
}
\label{tab:comparison}
\vspace{2mm}

\parbox{0.95\linewidth}{
\small
The perturbation magnitude $\epsilon$ is configured as 10\,cm for point perturbation and 50\,cm for point attachment attacks. For point detachment, 1\% of the points are removed from the input cloud.
}
\end{table}

\subsection{Results}
\begin{figure}[t]
    \centering
    \begin{minipage}{0.48\linewidth}
        \centering
        \includegraphics[width=\linewidth]{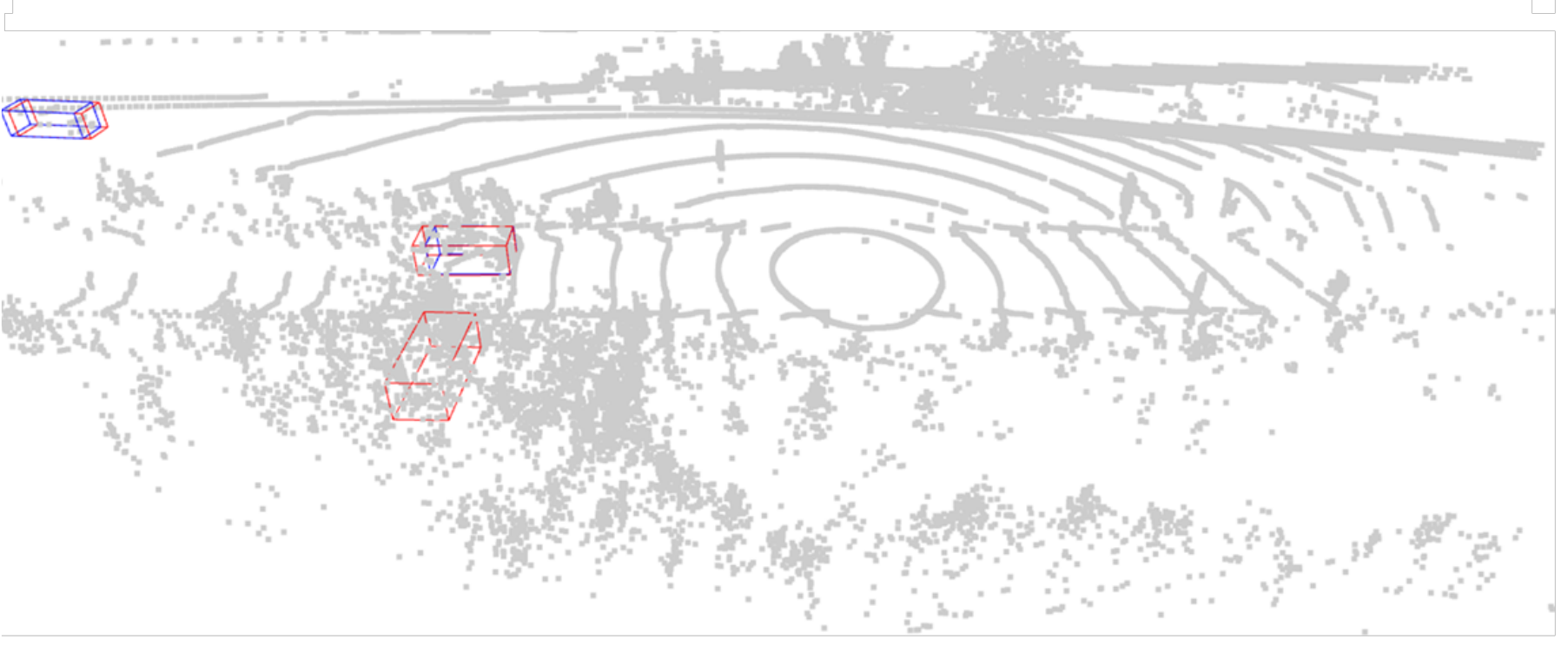}
        \vspace{2pt}
        (a)
    \end{minipage}
    \hfill
    \begin{minipage}{0.48\linewidth}
        \centering
        \includegraphics[width=\linewidth]{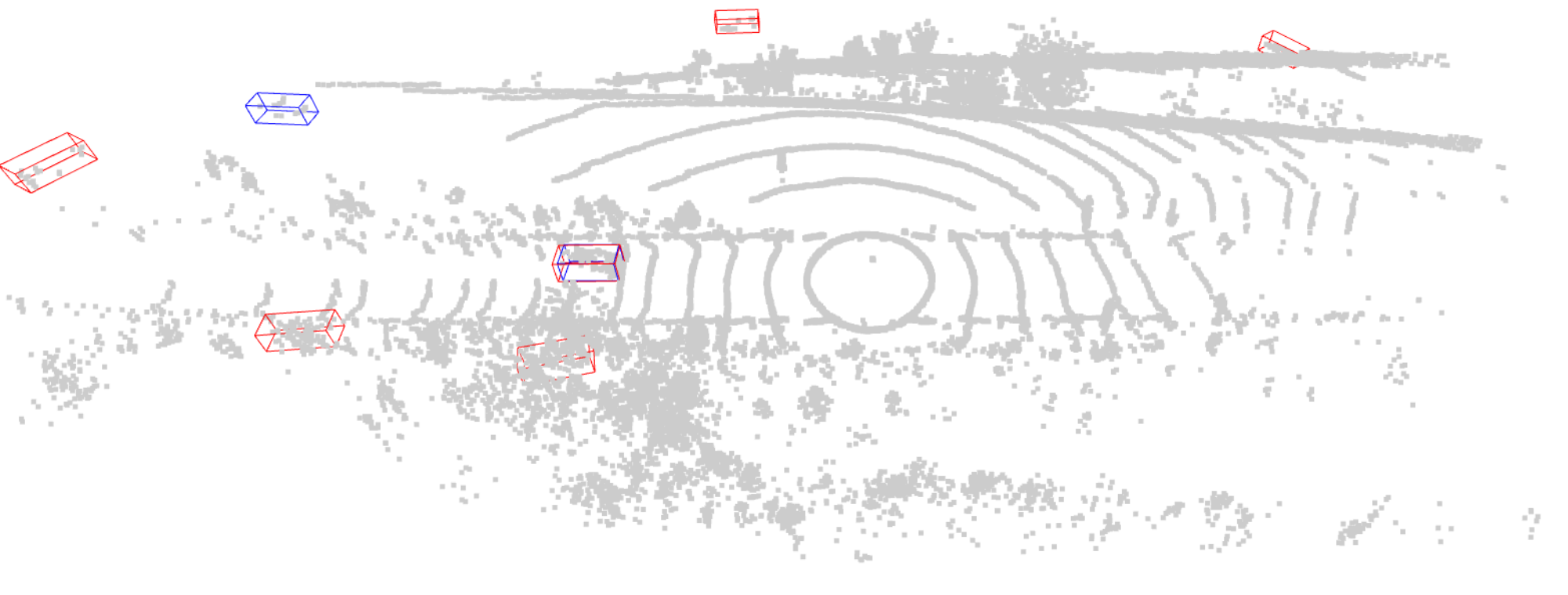}
        \vspace{2pt}
        (b)
    \end{minipage}
    
    \vspace{4pt}
    
    \begin{minipage}{0.48\linewidth}
        \centering
        \includegraphics[width=\linewidth]{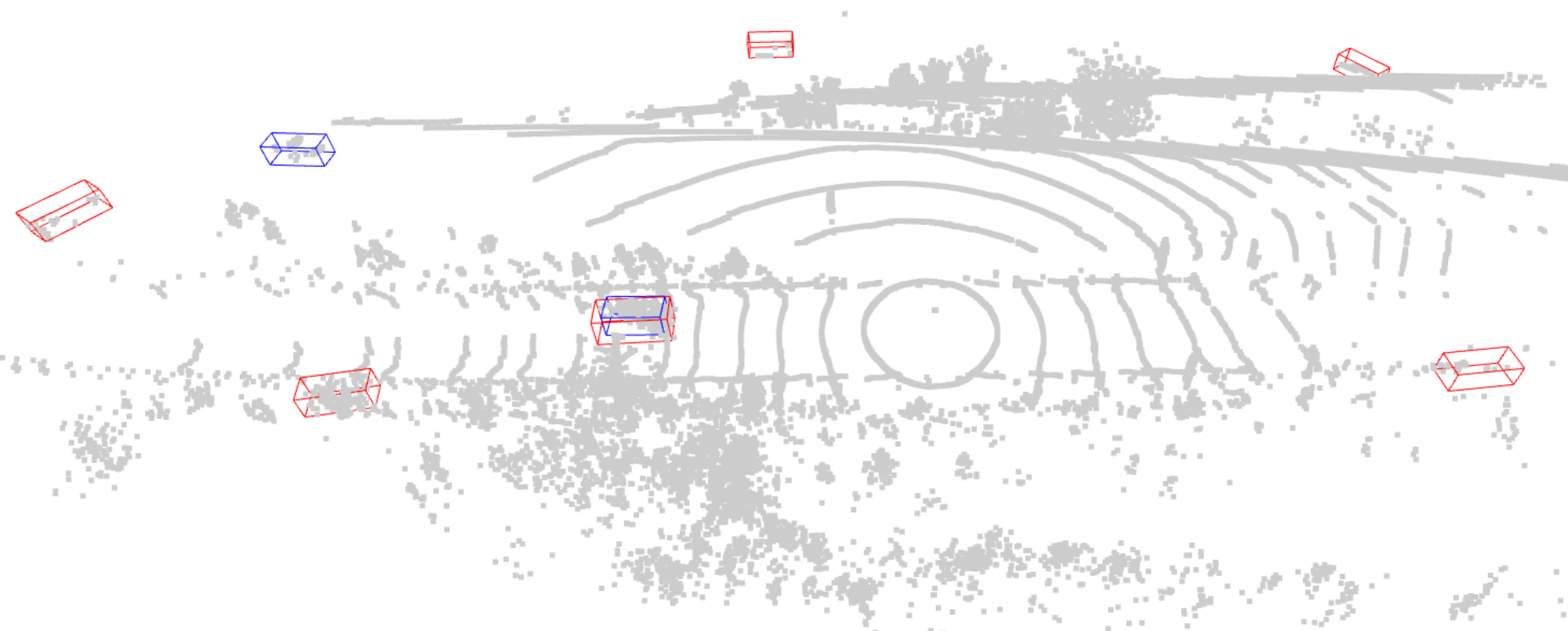}
        \vspace{2pt}
        (c)
    \end{minipage}
    \hfill
    \begin{minipage}{0.48\linewidth}
        \centering
        \includegraphics[width=\linewidth]{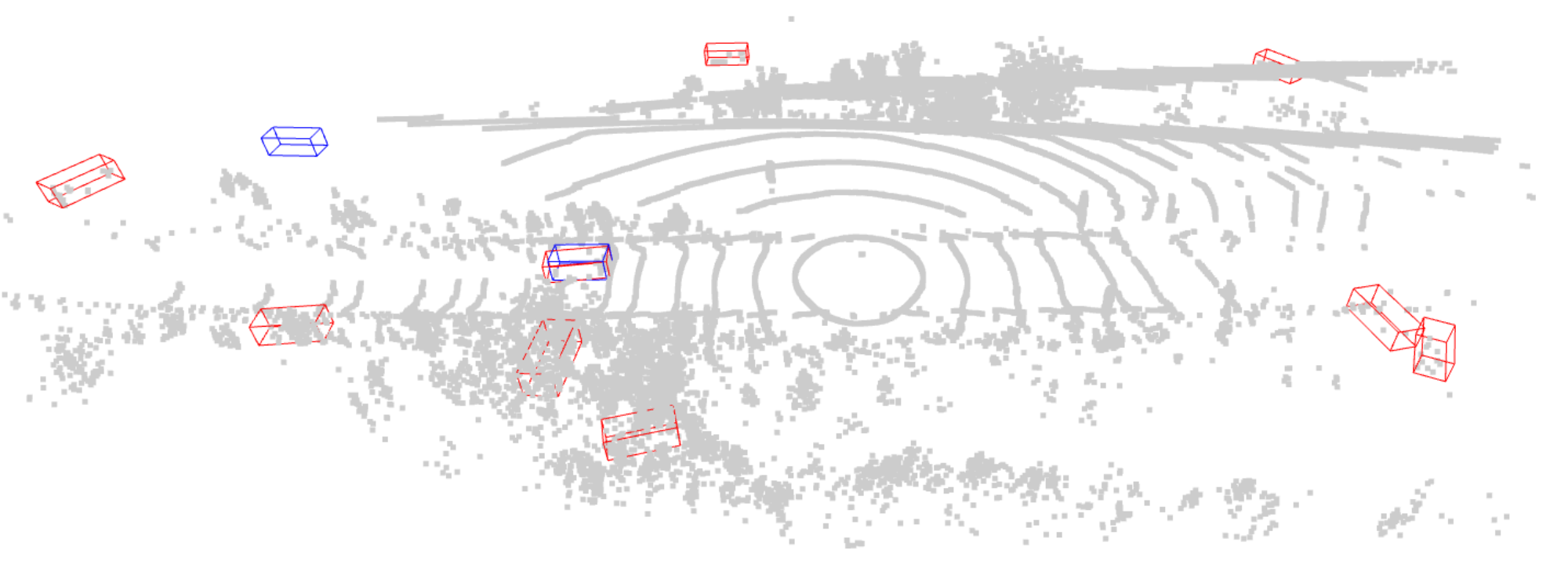}
        \vspace{2pt}
        (d)
    \end{minipage}

    \caption{Visualization of point clouds under various adversarial attack types. Ground truth 3D bounding boxes are depicted in blue, while predicted bounding boxes are shown in red. (a) Clean input point cloud. (b) Adversarially  perturbed input with $\epsilon = 0.1$.  (c) Point cloud with 300 adversarially attached points ($\epsilon = 0.5$). (d) Point cloud after removal of 1\% points.}
    \label{fig:4grid_figures}
\end{figure}

As shown in Table~\ref{tab:comparison}, the proposed simulation framework successfully generates adversarial LiDAR data that closely resemble the original point clouds while substantially reducing the detector’s performance. More specifically, point perturbation causes the greatest performance degradation with moderate perceptual distortion. Fig.~\ref{fig:4grid_figures} visually illustrates the impact of each adversarial attack on the point cloud, with the resulting degradation in detection performance evidenced by additional false positives and missed object detections. Notably, in the point removal case (Fig. ~\ref{fig:4grid_figures}d), although only a small fraction of points is dropped, the attack targets salient points, effectively erasing the object from the scene and leading to a complete detection failure.  Table~\ref{tab:attacs} provides a detailed analysis of each adversarial strategy under varying intensities. Point perturbation attacks lead to a consistent decline in detection accuracy as the perturbation budget increases. Point attachment attacks yield the smallest perceptual change, with Chamfer Distances remaining very low even for large $\epsilon$. Despite their imperceptibility, attachment attacks are less effective than perturbation, reinforcing the insight that altering native geometry is more detrimental to the detector than simply adding points

\begin{table}[ht]
\centering
\caption{Adversarial impact and perceptibility under different attack types on the SECOND detector.}
\label{tab:attacs}
\resizebox{0.48\textwidth}{!}{%
\begin{tabular}{llcc}
\toprule
\textbf{Attack Type} & \textbf{Parameter} & \textbf{mAP Ratio (\%)} & \textbf{Chamfer Distance (CD)} \\
\midrule
\multirow{6}{*}{Point Perturbation ($\epsilon$ in cm)} 
& 0.5  & 99.85 & 0.00010 \\
& 1    & 98.77 & 0.00025 \\
& 3    & 93.28 & 0.00182 \\
& 5    & 89.07 & 0.00466 \\
& 7    & 82.79 & 0.00842 \\
& 10   & 73.20 & 0.01533 \\
\midrule
\multirow{6}{*}{Point Detachment (drop ratio \%)} 
& 0.05 & 99.88 & 0.00023 \\
& 0.1  & 97.44 & 0.00117 \\
& 0.3  & 96.71 & 0.00921 \\
& 0.5  & 95.28 & 0.02177 \\
& 1.0  & 91.77 & 0.07035 \\
& 1.5  & 88.75 & 0.13397 \\
\midrule
\multirow{6}{*}{Point Attachment ($\epsilon$ in m)} 
& 0.05 & 98.82 & 0.00005 \\
& 0.1  & 95.19 & 0.00011 \\
& 0.3  & 87.72 & 0.00054 \\
& 0.5  & 78.46 & 0.00126 \\
& 0.7  & 77.24 & 0.00226 \\
& 1.0  & 77.09 & 0.00429 \\
\bottomrule
\end{tabular}
}

\end{table}

\section{Conclusion}

In this work, we designed and implemented an open-source integrated simulation framework for generating adversarial attacks on the perception and communication layers of autonomous vehicles. While our primary focus was on 3D perception, the framework is fully extensible to support 2D attacks. Integrated with ROS 2, the system features a modular, pluggable architecture compatible with real-world AV software stacks. We validated its effectiveness through a representative use case, demonstrating practical impact under realistic simulation conditions. Future work includes extending 2D attack support and developing a perception evaluation platform that computes vulnerability scores and suggests mitigation strategies.

    \label{fig:4wide_figures}

\bibliographystyle{IEEEtran}
\bibliography{bibliography}

\end{document}